\documentclass[prd,aps,preprint,tightenlines,floats,epsfig,superscriptaddress]{revtex4}
\usepackage{epsfig}
\usepackage{amsmath}
\usepackage{graphicx}
\begin{document}
% Page numbers bottom-center
%\pagestyle{plain}

\title{Neutron Electric Dipole Moment Constraint \\ on Scale of Minimal Left-Right Symmetric Model}

\author{Fanrong Xu}
\affiliation{Institute of Theoretical Physics, Chinese Academy of
Sciences, Beijing 100190, China} \affiliation{Maryland Center for
Fundamental Physics and Department of Physics, University of
Maryland, College Park, Maryland 20742, USA}
\author{Haipeng An}
 \affiliation{Maryland Center for Fundamental Physics and Department of Physics, University of
Maryland, College Park, Maryland 20742, USA}
\author{Xiangdong Ji}
 \affiliation{Maryland Center for Fundamental Physics and Department of Physics, University of
Maryland, College Park, Maryland 20742, USA}
 \affiliation{Institute of Particle Physics and Cosmology,
 Department of Physics, Shanghai Jiao Tong University, Shanghai 200240, China}
\affiliation{Center
for High-Energy Physics and Institute of Theoretical Physics, Peking
University, Beijing 100871, China}

\date{\today}

%\preprint{\vbox{\hbox{UMD-PP-08-00X}}}

\begin{abstract}
Using an effective theory approach, we calculate the neutron
electric dipole moment (nEDM) in the minimal left-right symmetric
model with both explicit and spontaneous CP violations. We integrate
out heavy particles to obtain flavor-neutral CP-violating effective
Lagrangian. We run the Wilson coefficients from the electroweak
scale to the hadronic scale using one-loop renormalization group
equations. Using the state-of-the-art hadronic matrix elements, we
obtain the nEDM as a function of right-handed W-boson mass and
CP-violating parameters. We use the current limit on nEDM combined
with the kaon-decay parameter $\epsilon$ to provide the most
stringent constraint yet on the left-right symmetric scale
$ M_{W_R} > (10 \pm 3) $ TeV.

\end{abstract}
\maketitle %\bigskip
%\bigskip

\section{introduction}

The electric dipole moment (EDM) of an elementary particle has been
a subject of strong interest from both experimentalists and
theorists for over half a
century~\cite{Harris:2007fz,Pospelov:2005pr}. A non-vanishing
intrinsic EDM indicates violation of time-reversal (T), parity (P)
and charge-conjugation-parity (CP) invariance at fundamental level.
Although the standard model (SM) of particle physics predicts
non-vanishing EDM for leptons and quarks from both the CP phase in
the Cabbibo-Kobayashi-Moskawa (CKM) matrix and the quantum
chromodynamic (QCD) $\theta$-angle, it is widely believed that new
physics must exit, allowing EDMs at least competitive with or even
dominate the SM predictions~\cite{Pospelov:2005pr,Ibrahim:1998je}.
New CP-violating physics is needed to explain, for example, the
baryon number asymmetry observed in the universe
today~\cite{Amsler:2008zzb}. In this article, we are interested in
the EDM of a strongly-interacting bound state---the free neutron.
Because of its charge-neutrality, the neutron EDM (nEDM) is
relatively ``easier" to measure. The current bound is $2.9 \times
10^{-26}$ $e$\,cm~\cite{Baker:2006ts}, which is already very
constraining for new physics. The upcoming experiments will enhance
the current sensitivity by one to two orders of
magnitude~\cite{Ito:2007xd}, which either will rule out many of the
new physics models under investigation, or will provide the first
opportunity to reveal an intrinsic nEDM.

Critical to understanding the experimental data is a theoretical
nEDM calculation with controlled precision of non-perturbative
strong-interaction physics. The result is indispensable for
extracting or constraining the new interaction parameters. In the
past, many calculations have been made in the literature, and most
of which are done in the context of ad hoc models
\cite{He:1989xj,Beall:1981zq,Abel:2004te,Arnowitt:1990eh,Arnowitt:1990je,He:1992db,Khatsymovsky:1992yg}.
In one class of models, the quark EDM is first obtained, and the
neutron EDM is calculated through constituent quark models. In
another class of models, the T-odd pion-nucleon interaction vertices
are first derived and then hadron physics effect is calculated
through pion loops. The relationship among different contributions
is often unclear and confusing. Depending on different modelings,
there are often large uncertainties in the final result.

In this paper, we follow an effective theory approach to calculate
the nEDM in the left-right symmetric model
(LRSM)~\cite{Mohapatra:1974gc}. The model was motivated by the
hypothesis that parity is a perfect symmetry at high-energy, and is
broken spontaneously at low-energy due to the asymmetric vacuum.
This model has a number of attractive features, including a natural
explanation of weak hyper-change in terms of baryon and lepton
numbers, existence of right-handed neutrinos and the seesaw
mechanism for neutrino masses, and possibility of spontaneous CP
violation. In a recent paper, we found a complete solution of the CP
violation structure of the minimal left-right symmetric model
(mLRSM)~\cite{Zhang:2007fn}. Our goal here is to derive a
factorization formula for nEDM in this model, with QCD and other
short-distance physics in the Wilson coefficients, and with
long-distance physics in hadronic matrix elements ready for, for
example, lattice QCD calculations. Using the state-of-the-art
hadronic matrix elements, we derive the best constraints on the
model parameters. In particular, we find the most stringent bound
yet on the left-right symmetric scale ~ $10\pm 3$ TeV, which is
beyond the detection capability of the Large Hadron Collider
(LHC)~\cite{Gninenko:2006br}.

Before starting, let us make a number of relevant comments. First of
all, it is possible that the entire nEDM to be measured can be
explained by the so-called QCD $\theta$-term, a term in the QCD
Lagrangian which will contribute to nEDM due to instanton effect.
Its contribution to nEDM has been calculated in several different
ways~\cite{Baluni:1978rf}. The nEDM constraint on $\theta$-term is
so strong $\theta<10^{-10}$~\cite{Baluni:1978rf} that there should
be some mechanism, for example the Peccei-Quinn
symmetry~\cite{Peccei:1977hh}, to cure this so-called strong CP
problem. We will not consider the induced $\theta$-contribution
to nEDM in mLRSM. However, this term seems unlikely to be the only or the
most important source for nEDM. A typical beyond-SM physics model
allow natural sizes of the nEDM on the order currently been probed by
experiments. Second, there has been a number of papers in the
literature about the perturbative QCD effects for nEDM in various
versions of LRSM without general CP
structure~\cite{Frere:1991jt,He:1992db,Shifman:1976de}. We will use
some of these results to make a coherent formulation, taking into
account various effects consistently. Finally, the EDM for spin
$\vec{s}$ has an interaction term in the hamiltonian $H = -
d\vec{s}\cdot\vec{E}/|\vec{s}|$, which corresponds to the following
term in the effective lagrangian density
\begin{equation}
 {\cal L} = -\frac{1}{2} d^E {\overline \psi}\sigma_{\mu\nu}i\gamma_5 \psi
 F^{\mu\nu}\ ,
\end{equation}
where $\sigma_{\mu\nu}=\frac{i}{2}[\gamma_\mu,\gamma_\nu]$ and
$F_{\mu\nu}$ is the electromagnetic field strength and $\psi$ is the
spin-1/2 Dirac field.

The presentation of the paper is organized as follows: In Sec. II,
we collect all the P-odd and CP-odd operators up to dimension-six
and discuss their short-distance QCD effects. In Sec. III, we
calculate the Wilson coefficients of quark EDM, CDM operators,
four-quark operators and the Weinberg operator in the framework of
mLRSM using effective theory by integrating out heavy particles. In
Sec. IV, we study nEDM in mLRSM numerically and get the lower bound
of the mass of the righthanded $W$-boson ($W_R$). It turns out that
the biggest contribution comes from four-quark operators. This
property is true not only in mLRSM, but for a large category of
models, for example two-Higgs-doublet Models (see
Ref.~\cite{Bigi:2000yz} for a good review). Recently, we have made a
comprehensive study of the matrix elements of the CP-odd and P-odd
four-quark operators in the neutron state~\cite{An:2009zh}, which
makes it possible to calculate the nEDM more reliably. We conclude
the paper in Sec. V.

\section{General CP-Violating Effective Lagrangian}

In this section, we lay out a general approach to calculating the
neutron EDM using the effective Lagrangian method, independent of
new physics. In this approach, one integrates out all heavy
particles including SM gauge bosons and heavy-quarks. The resulting
flavor neutral CP-violating effective Lagrangian has an expansion in
terms of operators consisting of light-quark fields, $u$, $d$, and
$s$ and the gluon field $G^{\mu\nu}$, with increasing dimensions,
\begin{equation}
   {\cal L}^{\rm CP-odd} = \sum_{i=4,...} {\cal L}^{\rm CP-odd}_i \ ,
\end{equation}
where ${\cal L}_i$ contains $i$-dimensional operator. The Wilson
coefficient of each operator depends on the renormalization scale
$\mu$ which in the end will be chosen as hadronic physics scale,
about 1 GeV or lattice cut-off $1/a$, where $a$ is lattice spacing.
The QCD operators also depend on the renormalization scale, but
physical results do not.

At the lowest dimension, there are two CP-odd operators,
\begin{equation}
    {\cal L}_4 = -\frac{g^2\theta}{32\pi^2} G^{\mu\nu}\tilde G_{\mu\nu} + \sum_q \tilde m_q \bar q
i\gamma_5 q\ ,
\end{equation}
Through $SU(3)$ chiral rotations, the CP-odd quark-mass term can be
rotated into a chiral singlet. Furthermore, one can eliminate either
$G\tilde G$ or the singlet quark-mass term through $U_A(1)$ chiral
rotation $q \rightarrow e^{i\gamma_5\alpha} q$.

At dimension-five level, there are two kinds of flavor neutral P-odd and
CP-odd operators, namely, the quark EDM operators and the chromo
electric dipole moment (CDM) operators,
\begin{equation}
{\cal L}_5=\sum _q d_q^E(\mu)O_q^E(\mu)+\sum_q d_q^C(\mu)O_q^C(\mu)
\ ,
\end{equation}
where $O_q^E = -\frac{1}{2} {\overline q}\sigma^{\mu\nu}i\gamma_5 q
F_{\mu\nu}$ and $ O_q^C = - \frac{1}{2}
 {\overline q}\sigma^{\mu\nu}i\gamma_5 t^a q G^a_{\mu\nu}$, and
$F_{\mu\nu}$ and $G^a_{\mu\nu}$ are the electromagnetic and gluon
field strengths, respectively, and $t^a$ are generators of the SU(3)
gauge group.  The one-loop evolution equations are
\cite{Frere:1991jt}
\begin{eqnarray}
\mu^2\frac{d}{d\mu^2} O_q^C(\mu) &=&
-\left(\frac{2}{3}-\frac{b_f}{2}\right)\frac{\alpha_S(\mu)}{4\pi}O_q^C(\mu) \ , \\
\mu^2\frac{d}{d\mu^2} O_q^E(\mu) &=&
-\frac{4}{3}\frac{\alpha_S(\mu)}{4\pi}O_q^E(\mu) \ ,
\end{eqnarray}
where $b_f=11-2n_f/3$, $n_f$ is the number of quark flavors. It is
easy to see that the dependence of the evolution of the quark CDM on
$n_f$ is the same as that of the strong coupling, since they are
both derived from wave function renormalization of the gluon field.

At dimension-six, there are a number of four-quark flavor-neutral
CP-odd operators and the Weinberg's three-gluon operator
\cite{Weinberg:1989dx},
\begin{equation}
 {\cal L}_6 = \sum_i C_i(\mu) O_{4i}(\mu) + C_g(\mu) O_g(\mu) \ ,
\end{equation}
where the four-quark CP-odd operators can be divided into two groups.
The first group includes operators with two different light flavors
\cite{Khatsimovsky:1987fr}
\begin{eqnarray}\label{4quark}
O_{11} &=& (\bar q i\gamma_5 q)(\bar q' q')\ , \nonumber\\
O_{12} &=& (\bar q q)(\bar q' i\gamma_5 q')\ , \nonumber\\
O_{21} &=& (\bar q i\gamma_5 t^a q)(\bar q' t^a q')\ , \nonumber\\
O_{22} &=& (\bar q t^a q)(\bar q' i\gamma_5 t^a q')\ , \nonumber\\
O_{3} &=& (\bar q i\gamma_5 \sigma^{\mu\nu} q)(\bar q'
\sigma_{\mu\nu} q') \ , \nonumber\\
O_{4} &=& (\bar q i\gamma_5 \sigma^{\mu\nu} t^a q)(\bar q'
\sigma_{\mu\nu} t^a q') \ ,
\end{eqnarray}
where $q, q'=u,d,s$ and $q\neq q'$. The second group includes
operators with one quark flavor
\begin{eqnarray}\label{4quark2}
O'_1 &=& (\bar q i\gamma_5 q)(\bar q q)\ , \nonumber\\
O'_2 &=& (\bar q i\gamma_5 t^a q)(\bar q t^a q)\ .
\end{eqnarray}
The Weinberg operator is defined as
\begin{equation}
O_g=-\frac{1}{6}f^{abc}\epsilon^{\mu\nu\alpha\beta}G^a_{\mu\rho}G^{b\rho}_{\nu}G^c_{\alpha\beta} \ ,
\end{equation}
where $\epsilon^{0123}=1$.

It is not difficult to see that all the dimension-six operators
listed above are CP-odd. In the first group there are two different
flavors in each operator. These operators are constructed by a
pseudoscalar current coupled to a scalar one, a pseudo-tensor
current coupled to a tensor one. No operator is constructed from an
axial-vector current coupled to a vector current since CP-odd
operators cannot be generated in this way. And since
$\gamma_5\sigma^{\mu\nu}=\frac{i}{2}\epsilon^{\mu\nu\alpha\beta}\sigma_{\alpha\beta}$,
the two operators $O_3$ and $O_4$ are enough to describe the CP-odd
pseudo-tensor and tensor coupling. Therefore, the first group
includes all the P and CP-odd four-quark operators constructed by
two different quark flavors. For the one-flavor case, the first four
operators in the first group automatically become $O'_1$ and $O'_2$
in the second group, and using the Fierz transformation one can
easily see that the operators described the tensor-pseudotensor
coupling are not independent of $O'_1$ and $O'_2$. Therefore, the
second group includes all the flavor-neutral CP-odd four-quark
operators with single quark flavor.

The leading-order QCD evolution equations for dimension-six operator
are as follows
\begin{eqnarray}\label{running}
\mu^2\frac{d}{d\mu^2}\left(\begin{array}{c}O_{11}\\O_{12}\\O_{21}\\O_{22}\\O_{3}\\O_{4}\\ \end{array}\right)&=&\frac{\alpha_S(\mu)}{4\pi}\left(\begin{array}{cccccc}8&0&0&0&0&1\\0&8&0&0&0&1\\0&0&-1&0&\frac{2}{9}&\frac{5}{12}\\
0&0&0&-1&\frac{2}{9}&\frac{5}{12}\\0&0&24&24&-\frac{8}{3}&0 \\
\frac{16}{3}&\frac{16}{3}&10&10&0&\frac{19}{3}\\ \end{array}\right)\left(\begin{array}{c}O_{11}\\O_{12}\\O_{21}\\O_{22}\\O_{3}\\O_{4}\\
\end{array}\right) \ , \\
\mu^2\frac{d}{d\mu^2}\left(\begin{array}{c}O'_{1}\\O'_{2}\\
\end{array}\right)&=&\frac{\alpha_S}{4\pi}\left(\begin{array}{cc}\frac{40}{9}&-\frac{4}{3}\\-\frac{80}{27}&-\frac{46}{9}\\
\end{array}\right)\left(\begin{array}{c}O'_{1}\\O'_{2}\\
\end{array}\right) \ , \\
\mu^2\frac{d}{d\mu^2}O_g&=&\frac{\alpha_S}{4\pi}\gamma_{gg}O_g \ .
\end{eqnarray}
The anomalous dimension of the Weinberg, $\gamma_{gg}$, has been
calculated in the literature~\cite{Braaten:1990gq},
$\gamma_{gg}=-C_A/2-n_f$, where $C_A=3$. The dimension-six operators
mix with the dimension-five operators when scale evolves, however at
the energy scale where only the light quarks exist, the mixing can
be neglected because the dimension-five quark EDM and CDM are
chirality flipping and thus proportional to the quark mass. At
higher energies, the mixing is important and we will discuss it in
the following sections.

There is no mixing between the Weinberg operator and the four-quark
operators listed in Eqs. (\ref{4quark}) and (\ref{4quark2}). To see
this, we can decompose the four-quark operators into irreducible
representations of the $SU(3)_L\times SU(3)_R$ chiral group and only
$(3,\bar 3)$, $(6,\bar 6)$, $(8,8)$ and their conjugate
representations are found~\cite{An:2009zh}. On the other hand, the
three-gluon operator is a chiral singlet. QCD evolution maintains
the chiral structure of operators.

When scale changes, the pure quark-gluon CP-odd operators generate perturbative contributions
to quark EDM through the following $T$-product
\begin{equation}
\int d^4x ~{\rm T}\left(eA_\mu(x)j^\mu_{em}(x){\sum_{i}}
'O_i(0)\right) \ ,
\end{equation}
where the summation neglects the quark EDM operator itself. The
contributions are divergent so they induce additional running of the
CP-odd operators. The contributions from the dimension-six operators
are proportional to the mass of light quarks and can be neglected.
The only large contribution is from the quark CDM operator, whose
running has an effective inhomogenous term, \cite{Frere:1991jt}
\begin{equation}
\mu^2\frac{d}{d\mu^2}
O_q^C=\frac{\alpha_S(\mu)}{4\pi}\left(-\left(\frac{2}{3}-\frac{b_f}{2}\right)O_q^C-\frac{16}{3}\frac{e}{g_S(\mu)}Q_qO_q^E\right)
\ ,
\end{equation}
where $Q_q$ is the electric charge of the quarks and $g_S$ is the
coupling of strong interaction. Inversely, the quark EDM operators
can also generate quark CDM operators through the electromagnetic
interaction which is, however, proportional to the electromagnetic
fine-structure constant.

Therefore, omitting the $\theta$-contribution, one can define the
following electric dipole form factor
\begin{eqnarray}\label{fac}
&&-F_n^E(q^2) \bar U_n(\vec k_2) \sigma^{\mu\nu}\gamma_5 q_\mu U_n(\vec k_1)\epsilon_\nu(q)\nonumber\\
&=&\langle N(\vec k_2)|\sum_q d_q^E(\mu)O_q^E(0;\mu)\nonumber \\
&&+ i\int d^4x\;{\rm T}\left[eA_\mu(x) j_{em}^\mu(x) \left(\sum_q
d_q^C(\mu)O^C_q(0;\mu) \right.\right.\nonumber\\
&&\left.\left.+\sum_i
C_i(\mu)O_{4i}(0;\mu)+C_g(\mu)O_g(0;\mu)\right)\right]|\gamma(q)N(\vec
k_1)\rangle \ ,
\end{eqnarray}
where $q^\mu = k_2^\mu - k_1^\mu$ and $U_n$ is the wavefunction of
neutron and $\epsilon^\nu$ is the polarization of the incoming
photon. The static nEDM is just the zero-momentum limit of the form
factor $d^E_n = F^E_n(0) $.

%In LRSM, not all the four-quark operators in (\ref{4quark}) appear
%in the leading order. What appear are the combinations
%$O_{11}-O_{12}$ and $O_{21}-O_{22}$ as will be discuss in detail in
%Sec. IV.

\section{Wilson Coefficients in LRSM}

Following the previous section, we make calculation of nEDM in the
mLRSM by first evaluating the Wilson coefficients of the effective
quark-gluon operators at the electroweak scale, and subsequently
running them to hadronic scale. The detail of the model can be found
in Ref.~\cite{Zhang:2007fn}, in which the spontaneous CP-violation
is controlled by a phase angle $\alpha$ in the Higgs sector, and
additional parameters of the model include, among others, the masses
of the right-handed gauge boson and the new Higgs bosons. In the
following subsections, we study the Wilson coefficients of various
CP-violating operators separately. We will ignore the contribution
of the $\theta$-term as it will usually generate a much too large
nEDM: We assume certain mechanisms such as Peccei-Quinn
symmetry~\cite{Peccei:1977hh} is in operation to suppress it.

\subsection{CP-Odd Four-Quark Operators}

\begin{figure}[b]
\begin{center}
\includegraphics[width=6cm]{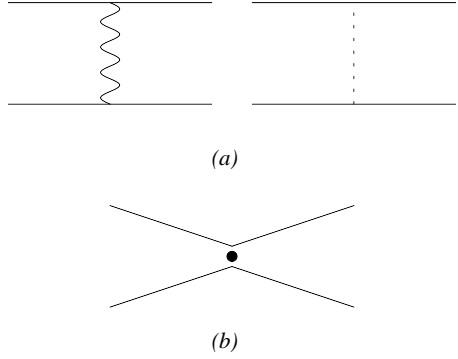}
\caption{Effective four-quark operators generated by integrating out
$W_1$-boson: (a) the diagrams in the full theory and (b) the
effective operator. }\label{fourquark}
\end{center}
\end{figure}

To leading order, diagrams in Fig. \ref{fourquark} generate the
CP-odd four-quark operators induced by the exchange of gauge bosons
and Higgs bosons. The operators are listed in Eq. (\ref{4quark}) and
(\ref{4quark2}). The corresponding Wilson coefficients can be easily
read through the diagrams,
\begin{eqnarray}
C_{11}^{ab}&=&\frac{\sqrt{8}G_F}{6}\sin2\zeta{\rm
Im}(e^{-i\alpha}V_L^{ab}V_R^{ab*})+\frac{\sqrt{8}G_F}{M_{H_0}^2}{\rm
Im}(C^{aa}D^{bb})\nonumber\\
&&+\frac{\sqrt{8}G_F}{6M_{H_2}^2}(m_a^2-m_b^2)\xi\;{\rm Im}(e^{-i\alpha}V_L^{ab}V_R^{ab*}) \ , \nonumber\\
C_{12}^{ab}&=&-\frac{\sqrt{8}G_F}{6}\sin2\zeta{\rm
Im}(e^{-i\alpha}V_L^{ab}V_R^{ab*})+\frac{\sqrt{8}G_F}{M_{H_0}^2}{\rm
Im}(C^{aa}D^{bb})\nonumber\\
&&+\frac{\sqrt{8}G_F}{6M_{H_2}^2}(m_a^2-m_b^2)\xi\;{\rm Im}(e^{-i\alpha}V_L^{ab}V_R^{ab*})\ , \nonumber\\
C_{21}^{ab}&=&{\sqrt{8}G_F}\sin2\zeta{\rm
Im}(e^{-i\alpha}V_L^{ab}V_R^{ab*})\nonumber\\
&&+\frac{\sqrt{8}G_F}{M_{H_2}^2}(m_a^2-m_b^2)\xi\;{\rm Im}(e^{-i\alpha}V_L^{\alpha\beta}V_R^{\alpha\beta*})\ , \nonumber\\
C_{22}^{ab}&=&-{\sqrt{8}G_F}\sin2\zeta{\rm
Im}(e^{-i\alpha}V_L^{ab}V_R^{ab*})\nonumber\\
&&+\frac{\sqrt{8}G_F}{M_{H_2}^2}(m_a^2-m_b^2)\xi\;{\rm
Im}(e^{-i\alpha}V_L^{\alpha\beta}V_R^{\alpha\beta*})\ ,
\end{eqnarray}
\begin{eqnarray}
C_3^{ab}&=&\frac{\sqrt{8}G_F}{6M_{H_2}^2}(m_a^2-m_b^2)\xi\;{\rm Im}(e^{-i\alpha}V_L^{\alpha\beta}V_R^{\alpha\beta*})\ , \nonumber\\
C_4^{ab}&=&\frac{\sqrt{8}G_F}{M_{H_2}^2}(m_a^2-m_b^2)\xi\;{\rm Im}(e^{-i\alpha}V_L^{\alpha\beta}V_R^{\alpha\beta*})\ , \nonumber\\
C_{11}^{aa'}&=&\frac{2\sqrt{8}G_F}{M_{H_0}^2}{\rm
Im}(C^{aa}C^{a'a'*})\ , \nonumber\\
C_{12}^{aa'}&=&-\frac{2\sqrt{8}G_F}{M_{H_0}^2}{\rm
Im}(C^{aa}C^{a'a'*})\ , \nonumber\\
C_{11}^{bb'}&=&\frac{2\sqrt{8}G_F}{M_{H_0}^2}{\rm
Im}(D^{aa}D^{a'a'*})\ , \nonumber\\
C_{12}^{bb'}&=&-\frac{2\sqrt{8}G_F}{M_{H_0}^2}{\rm
Im}(D^{aa}D^{a'a'*})\ ,
\end{eqnarray}
where $a,a'\in{u,c,t}$, $a\neq a'$ and $b,b'\in{d,s,b}$, $b\neq b'$,
$C=V_L\hat M_D V_R^\dagger-2\xi e^{i\alpha}\hat M_U$,
$D=V_L^\dagger\hat M_U V_R-2\xi e^{-i\alpha} \hat M_D$, $M_{H_0}$ is
the mass of the flavor changing neutral Higgs (FCNH) and $M_{H_2}$
is the mass of $H_2^+$ which is a charged Higgs in
mLRSM~\cite{Zhang:2007fn}. $\hat M_U$ and $\hat M_D$ are
diagonalized quark mass matrices. $\zeta$ is the mixing angle
between the lefthanded and righthanded $W$-bosons that
\begin{equation}
\sin2\zeta\simeq -r\frac{4m_b}{m_t}\left(\frac{M_1}{M_2}\right)^2 \
,
\end{equation}
where $r\equiv(m_t/m_b)\xi$ and $\xi$ is the ratio between the two
vevs of the Higgs bidoublet in mLRSM~\cite{Zhang:2007fn}. The
contributions due to the Higgs exchanges are always proportional to
quark masses. Since we are only interested in operators with at
least two of the quarks being light, the Wilson coefficients are
always proportional to at least one light quark mass, or they are
proportional to heavy quark masses but must be suppressed by the
non-diagonal CKM matrix elements. Furthermore, the mass of FCNH is
strongly constrained to very large value by the mass differences and
the CP-violating decay properties of the neutral K-bosons and
B-bosons~\cite{Zhang:2007fn,Kiers:2002cz}, and detailed calculation
shows $H_2^+$ is as heavy as FCNH. If we are interested in the case
of a few TeV right-handed W-boson mass, we can safely neglect the
Higgs exchange contributions. Then at the electroweak scale the
Wilson coefficients of the CP-odd four-quark operators can be
simplified to
\begin{eqnarray}\label{wilson4quark}
C_{11}^{ab}&=&-C_{12}^{ab}=\frac{\sqrt{8}G_F}{6}\sin2\zeta\;{\rm
Im}(e^{-i\alpha}V^{ab}_LV^{ab*}_R) \ , \nonumber\\
C_{21}^{ab}&=&-C_{22}^{ab}={\sqrt{8}G_F}\sin2\zeta\;{\rm
Im}(e^{-i\alpha}V^{ab}_LV^{ab*}_R) \ .
\end{eqnarray}
We will take this simple limit in the following discussion.

\subsection{Quark EDM and CDM Operators}

\begin{figure}[hbt]
\begin{center}
\includegraphics[width=7cm]{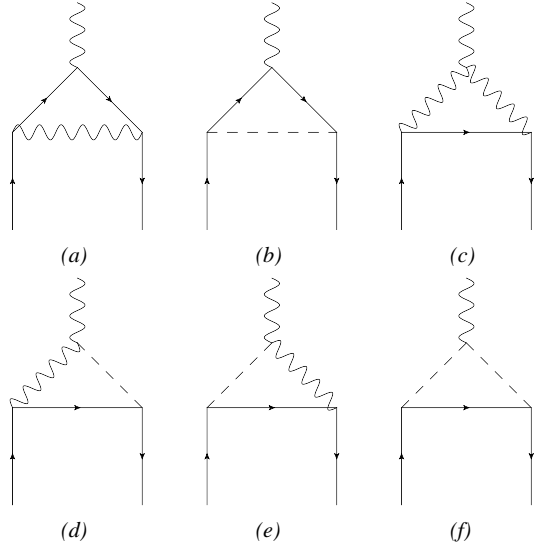}
\caption{One-loop contribution to quark EDM. The internal wavy lines represent
the W-boson contribution and the dashed lines the corresponding Goldstone bosons.} \label{soni}
\end{center}
\end{figure}

The one-loop contributions to the quark EDM from the gauge
interactions are shown in Fig. \ref{soni}, where the internal wavy
lines represent the light charged gauge-boson $W_1$ which is
dominated by $W_L$, but has a small admixture of $W_R$. The dashed
lines represent the charged-Goldstone boson present in Feynman
gauge, and the external wavy line is the static electric field or
photon. Diagrams a) and b) have the photon interacting with the
quarks directly, and these from c) to f) have the photon interacting
with charged bosons. For the quark CDM case we have the first two
diagrams only with the external wavy line representing a gluon.

These diagrams have been calculated in the literature long ago
\cite{Beall:1981zq}, our result is somewhat different from theirs in
the infrared part.
The CP-odd part of the diagrams in Fig. \ref{soni} can be expressed
in terms of the coefficients of the EDM and CDM operators. For the
up quark, we have $d_u^E O_u^E + d_u^C O_u^C$ with,
\begin{eqnarray}\label{udm}
d_{u}^E&=&\frac{1}{16\pi^2}\sum_{i=d,s,b}m_{di}e\sqrt{8}G_F\sin2\zeta{\rm
Im}(e^{-i\alpha}V_L^{1i}V_R^{1i*})\nonumber\\
&\times&\frac{1}{(1-r_i)^3}\left(\frac{4}{3}-4r_i+3r_i^2-\frac{1}{3}r_i^3+\frac{1}{2}r_i\ln
r_i-\frac{3}{2}r_i^2\ln r_i\right) \ , \nonumber\\
d_u^C&=&\frac{1}{16\pi^2}\sum_{i=d,s,b}m_{di}g_s\sqrt{8}G_F\sin2\zeta{\rm
Im}(e^{-i\alpha}V_L^{1i}V_R^{1i*})\nonumber\\
&\times&\frac{1}{(1-r_i)^3}\left(1-\frac{3}{4}r_i-\frac{1}{4}r_i^3+\frac{3}{2}r_i\ln
r_i\right) \ .
\end{eqnarray}
And for the down quark, the contribution is $d_d^E O_d^E + d_d^C
O_d^C$ with
\begin{eqnarray}\label{ddm}
d_d^E&=&\frac{1}{16\pi^2}\sum_{i=u,c,t}m_{ui}e\sqrt{8}G_F\sin2\zeta{\rm
Im}(e^{-i\alpha}V_L^{1i}V_R^{1i*})\nonumber\\
&\times&\frac{1}{(1-r_i')^3}\left(\frac{5}{3}-\frac{17}{4}r'_i+3{r'_i}^2-\frac{5}{12}{r'_i}^3+r'_i\ln
r'_i-\frac{3}{2}{r'_i}^2\ln r'_i\right)\ , \nonumber\\
d_d^C&=&-\frac{1}{16\pi^2}\sum_{i=u,c,t}m_{ui}g_s\sqrt{8}G_F\sin2\zeta{\rm
Im}(e^{-i\alpha}V_L^{i1}V_R^{i1*})\nonumber\\
&\times&\frac{1}{(1-r'_i)^3}\left(1-\frac{3}{4}r'_i-\frac{1}{4}{r'_i}^3+\frac{3}{2}r'_i\ln
r'_i\right) \ .
\end{eqnarray}
It is easy to see that this mixing angle is suppressed by the mass
ratio of bottom and top quarks and by the ratio of the left and
right handed $W$-boson masses. $m_{ui}$ are the masses of up-type
intermediate quarks, $r_i =m_{di}^2/M_W^2$, $r'_i=m_{ui}^2/M_W^2$,
$V_L$ and $V_R$ are the left and right-handed CKM mixing matrices,
respetively, $\alpha$ is the spontaneous CP phase mentioned earlier.

% To be specific, let us consider the dipole moment of a down quark.
%The intermediate quark flavors contributing to the EDM include up,
%charm, and top quarks, separately. The contribution from diagrams 1)
%and 2) is easy to calculate and can be expressed in terms the
%dipole-momentum operator $-(1/2)\Delta_1d_d\bar d \sigma^{\mu\nu}i
%\gamma_5 F_{\mu\nu}d$ with a coefficient
%\begin{equation}
%\Delta_{1}d_d=-\frac{1}{32\pi^2}g_Lg_R\sin
%\zeta\sum_i\frac{m_{ui}e}{M_W^2}{\rm
%Im}(e^{-i\alpha}V_L^{1i}V_R^{1i*})\frac{1}{(1-r_i^3)}\left(\frac{2}{3}-\frac{1}{2}r_i-\frac{1}{6}r^3_i+r_i\ln
%r_i\right) \label{quarkd}
%\end{equation}
%And for down quark

\begin{figure}[hbt]
\begin{center}
\includegraphics[width=6cm]{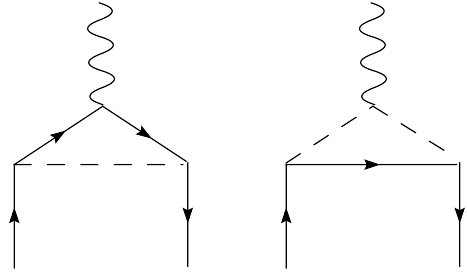}
\caption{Higgs-induced quark EDM. The dashed lines here represents the Higgs bosons.} \label{sonihiggs}
\end{center}
\end{figure}

In mLRSM, $H_2^+$ also gives contribution to the quark EDM and CDM.
The relevant diagrams are shown in Fig. \ref{sonihiggs}, and the
result is
\begin{eqnarray}\label{Hdm}
d^E_u&=&-\sum_{a\in\{d,s,b\}}\frac{1}{16\pi^2}\sqrt{8}G_F\frac{2m_a(m_u^2-m_a^2)}{M_{H_2}^2}\xi\;{\rm Im}(e^{-i\alpha} V_L^{1a}V_R^{1a*})\nonumber\\
&&\left[e_d\frac{3-4r_j+r_j^2+2\ln
r_j}{2(-1+r_j)^3}-e\frac{-1+r_j^2-2r_j\ln
r_j}{2(-1+r_j)^3}\right]\;,\nonumber\\
d^C_u&=&\sum_{a\in\{d,s,b\}}\frac{g_s}{16\pi^2}\sqrt{8}G_F\frac{2m_a(m_u^2-m_a^2)}{M_{H_2}^2}{\rm\;
Im}(e^{-i\alpha} V_L^{1a}V_R^{1a*})\frac{3-4r_j+r_j^2+2\ln
r_j}{2(-1+r_j)^3}\;,
\end{eqnarray}
\begin{eqnarray}\label{Hdm1}
d^E_d&=&-\sum_{a\in\{u,c,t\}}\frac{1}{16\pi^2}\sqrt{8}G_F\frac{2m_a(m_a^2-m_d^2)}{M_{H_2}^2}\xi\;{\rm Im}(e^{-i\alpha} V_L^{a1}V_R^{a1*})\nonumber\\
&&\left[e_u\frac{3-4r_j'+r_j'^2+2\ln
r_j'}{2(-1+r_j')^3}+e\frac{-1+r_j'^2-2r_j'\ln
r_j'}{2(-1+r_j')^3}\right]\;,\nonumber\\
d^C_d&=&\sum_{a\in\{u,c,t\}}\frac{g_s}{16\pi^2}\sqrt{8}G_F\frac{2m_a(m_a^2-m_d^2)}{M_{H_2}^2}\xi\;{\rm
Im}(e^{-i\alpha} V_L^{a1}V_R^{a1*})\frac{3-4r_j'+r_j'^2+2\ln
r_j'}{2(-1+r_j')^3}\;,
\end{eqnarray}
in which
\begin{eqnarray}
r_j&=&\frac{M_{Dj}^2}{M_{H_2}^2}\;,\;\;r_j'\;=\;\frac{M_{Uj}^2}{M_{H_2}^2}\;,
\end{eqnarray}
Therefore, if the right-handed $W$-boson has a moderate mass, say, a
few TeV, the contribution from $H_2^+$ to the quark EDM and CDM can
be neglected in comparison to that from the right-handed gauge
boson.

%The mixing angle $\zeta\sim(M_1/M_2)^2\xi$.
%Therefore, compared CDM and EDM operators induced by the gauge
%bosons, the ones induced by the charged Higgs is suppressed by
%$m_q^2/M_1^2$. Therefore, the Higgs induced part can be safely
%neglected.
% where $g_L$ and $g_R$ are gauge coupling constants for left and
%right SU(2) groups, respectively, $\zeta$ is the mixing angle
%between the two charged W bosons, $m_{ui}$ the masses of up-type
%ntermediate quarks, $r_i =m_{ui}^2/M_W^2$, $V_L$ and $V_R$ the left
%and right-handed CKM mixing, respetively, $\alpha$ is the
%spontaneous CP phase mentioned earlier.
%This result is similar to that in Ref. xxx, except that there is no infrared divergence in the limit when the quark %mass goes to zero.

\begin{figure}[hbt]
\begin{center}
\includegraphics[width=6cm]{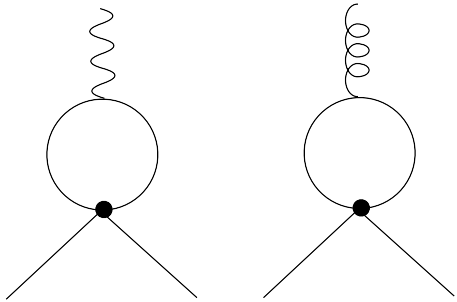}
\caption{Long-distance contributions to quark EDM and CDM through
CP-odd four-quark operators. }\label{longdis}
\end{center}
\end{figure}

Actually, there are both long-distance and short-distance
contributions from the one-loop diagrams in Fig. \ref{soni} and Fig.
\ref{sonihiggs}. The short-distance contributions come from the
integration region where the internal momentum is around $M_W$; and
the long-distance one from the loop momentum around the internal
light quark masses. Due to asymptotic freedom of the strong
interaction, the short-distance contributions can be calculated
accurately using perturbation theory. The long-distance
contributions, however, suffer from non-perturbative QCD effects,
and the only known way to calculate it correctly is by Lattice QCD.
In the matching calculation, the long distance contribution has to
be subtracted to obtain the Wilson coefficients, which is shown in
Fig. \ref{longdis}.
This contribution can be calculated using a certain UV regulator,
such as dimensional regulation or momentum cut-off or lattice
regularization. Any regularization preserving a certain Fierz
identity will give a zero answer as the loop integral involves only
the photon or gluon external momentum. Other regularizations, such
as naive dimensional regularization, will find a finite
contribution. One must be careful though that the vanishing of
long-distance contribution is only true at one-loop level: as soon
as the QCD corrections are taken into account, the result becomes
non-zero. Therefore, to the leading order, we can directly read off
the Wilson coefficients of quark EDM and CDM operators from
Eqs.(\ref{udm}), (\ref{ddm}), and (\ref{Hdm}).

\subsection{Weinberg Operator}

\begin{figure}[hbt]
\begin{center}
\includegraphics[width=12cm]{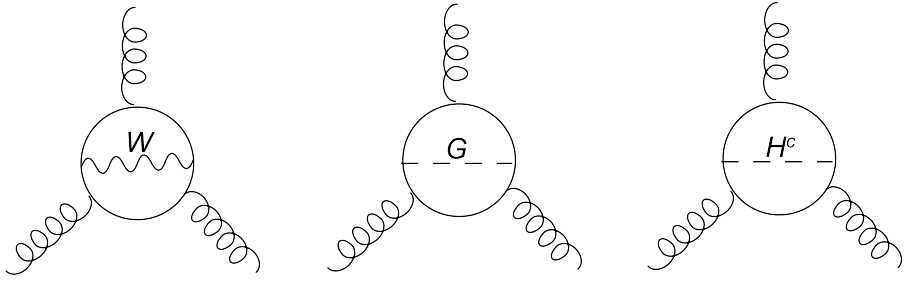}
\caption{Diagrams contributing to Weinberg operator in mLRSM. The
first diagram is induced by the W-boson exchange, the second by
Goldstone exchange and the third by the charged Higgs
boson.}\label{weinberg}
\end{center}
\end{figure}

In mLRSM, the Weinberg operator can be induced from diagrams in Fig.
\ref{weinberg}. Since the result is proportional to the quark
masses, the leading contribution comes from the third generation of
the quarks running in the loop. These are two-loop diagrams, the
Weinberg operator comes out after one integrates out the internal
quarks and bosons entirely. If one follows the effective theory
approach, in which the top quark and the W-boson are first
integrated out, the CDM operator of the bottom quark emerges and one
can get its wilson coefficient from Eq. (\ref{ddm}).

\begin{figure}[hbt]
\begin{center}
\includegraphics[width=4cm]{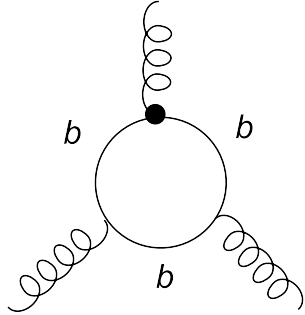}
\caption{Contribution to the three-gluon vertex after
integrating out the top quark, the Higgs boson and the W-bosons. The
black dot labels the bottom quark CDM operator.}\label{weinberg1}
\end{center}
\end{figure}

Then from Fig. \ref{weinberg1}, one gets the major contribution to
the Weinberg three gluon vertex. Because this diagram would diverge
quadratically in the infrared if the mass of the bottom quark was
zero, this diagram should be proportional to $1/m_b^2$. However,
chirality flipping is needed or otherwise the fermion loop will
vanish, so the numerator of the diagram must be proportional to
$m_b$. Combining the two effects together, this diagram is
proportional to $d^C_b/m_b$, where $d^C_b$ is the bottom quark CDM
which is proportional to $m_t$. Therefore this diagram has an
enhancement of a factor of $m_t/m_b$, about 40, which was first
found in Ref.~\cite{Chang:1990sfa}. Detailed calculation gives the
Wilson coefficient
\begin{equation}
C_g(m_b)=\frac{g_s^2(m_b)}{16\pi^2}\frac{d^C_b(m_b)}{m_b}\;.
\end{equation}
This contribution is seemingly large, however, it is suppressed by a
numerical factor, $1/(1-m_t^2/M_1^2)^3\simeq-0.02$ in Eq.
(\ref{ddm}). Therefore, the effect of the enhancement is totally
canceled. Furthermore, the evolution also makes the contribution of
this operator to be smaller at the low energy
region~\cite{Braaten:1990gq}. Therefore, we safely neglect its
contribution to nEDM in the following calculations.

\subsection{Wilson Coefficients at Hadronic Scale Through Leading-Order QCD Evolution}

The coefficient functions above, and hence the quark-gluon
operators, are calculated at the high-energy electroweak scale,
which is not yet useful for practical calculations. We are going to
remedy this by running down the scale in the composite operator by
including the leading logarithmic pQCD corrections. When we change
the scale, dimension-six operators will mix with each other and
generate dimension-five operators, and dimension-five operators will
also mix with each other. The Wilson coefficients for CP-odd
four-quark operators are shown to the leading order approximation in
Eq. (\ref{wilson4quark}). From Eq. (\ref{running}) the
renormalization group equations (RGE) keep this relation, and other
CP-odd four-quark operators are not generated by the running. Then
one can redefine the operators
\begin{eqnarray}
O_1^{ab}&=&O_{11}^{ab}-O_{12}^{ab}\ ,\nonumber\\
O_2^{ab}&=&O_{21}^{ab}-O_{22}^{ab}\ ,
\end{eqnarray}
with the Wilson coefficients $C_1^{ab}=C_{11}^{ab}$ and
$C_2^{ab}=C_{21}^{ab}$, respectively. Therefore, the RGEs of the
Wilson coefficients of the dimension-six operators can be written as
\begin{eqnarray}
\mu^2\frac{d}{d\mu^2}C_{1}^{ab}(\mu)&=&-8\frac{\alpha_s(\mu)}{4\pi}C_{1}^{ab}(\mu)\;;\nonumber\\
\mu^2\frac{d}{d\mu^2}C_{2}^{ab}(\mu)&=&\frac{\alpha_s(\mu)}{4\pi}C_{2}^{ab}(\mu)\;,
\end{eqnarray}
which shows that $C_1$ grows as the scale goes down, whereas $C_2$ does the opposite.

The RGE of the quark CDM operators are a little bit complicated. For
$d$ quark and $s$ quark CDM operators, as we discussed before, the
$c$ quark internal line gives a large contribution. Therefore, the
RGEs of $d$ and $s$ quark CDM operators can be written
as~\cite{Frere:1991jt}
\begin{eqnarray}
\mu^2\frac{d}{d\mu^2}d^{C}_{d,s}(\mu)&=&-\frac{g_s^3(\mu)}{(16\pi)^2}m_c(\mu)\left(\frac{2}{3}\gamma_{31}C_2^{c(d,s)}(\mu)-4\gamma_{32}C_1^{c(d,s)}(\mu)\right)\nonumber\\
&&-\frac{g_s^2(\mu)}{16\pi^2}(\gamma_{33}+b_f/2-\delta)d^C_{d,s}(\mu)\
.
\end{eqnarray}
The Wilson coefficient of the up quark CDM operator is one order of
magnitude smaller than that of the of d quark due to that
$m_s/m_c\sim1/10$. In the above formula, $\gamma_{31}=5/2$,
$\gamma_{32}=-1$, $\gamma_{33}=-14/3$, and $\delta=-4$ is the
anomalous dimension of the quark mass. Detailed calculation gives,
at $m_c$, the relevant Wilson coefficients are
\begin{eqnarray}
C_1^{u(d,s)}(m_c)&=&3.0C_1^{u(d,s)}(M_L)\ ,\nonumber\\
C_2^{u(d,s)}(m_c)&=&0.87C_2^{u(d,s)}(M_L)\ ,\nonumber\\
d^C_{d,s}(m_c)&=&1.7\frac{m_c}{16\pi^2}C_1^{c(d,s)}(M_L)+0.34\frac{m_c}{16\pi^2}C_2^{c(d,s)}(M_L)+1.6d^C_{d,s}(M_L)\
.
\end{eqnarray}
were $M_L$ is the mass of the SM W-boson.

The CP-odd operators generate additional running of the quark EDM
operators through the electromagnetic interaction. The RGE of the
down quark EDM operator can be written as~\cite{Frere:1991jt}
\begin{eqnarray}
\mu^2\frac{d}{d\mu^2}d^E_d(\mu)=-\frac{2}{3}\frac{em_c(\mu)g^2_s(\mu)}{(16\pi^2)^2}\gamma_{41}C_2^{cd}(\mu)-\frac{eg_s(\mu)}{16\pi^2}\gamma_{43}d^C_d(\mu)-\frac{g^2_s(\mu)}{16\pi^2}(\gamma_{44}-\delta)d^E_d(\mu)\
,
\end{eqnarray}
where $\gamma_{41}=16/3$, $\gamma_{43}=16/9$, $\gamma_{44}=-16/3$, and similarly for the strange quark.
The RGE of the electromagnetic coupling $e$ does not depend on the
strong coupling constant $g_s$ up to one-loop, therefore, can be
treated as a constant. At the charm quark mass scale, one can get
\begin{eqnarray}
d^E_{d,s}(m_c)&=&\frac{em_c}{16\pi^2}(0.07C^{c(d,s)}_1(M_L)+0.34C_2^{c(d,s)}(M_L))\nonumber\\
&&+0.17ed^C_{d,s}(M_L)+0.83d^E_{d,s}(M_L)\ .
\end{eqnarray}
which shows the explicit contributions from the running of the four-quark operators
as well as CDM operators.

\section{nEDM in mLRSM and Constraint on Left-right Symmetry scale}

In this section, we carry out the last step of the nEDM calculation in
mLRSM by incorporating the neutron matrix elements of hadronic
operators. We collect the state-of-art results in the literature and
use them to constrain the parameters in mLRSM. We find that in order to
satisfy the current experimental bound on nEDM and
the data on kaon-decay parameter $\epsilon$, the right-handed gauge
boson $W_R$ might be as heavy as $10\pm 3$ TeV. This bound is far higher
than the bound obtained previously from the kaon mass difference, making
it difficult to discover left-right symmetry at LHC.

\subsection{Hadronic Matrix Elements}

The most difficult part in calculating nEDM is to estimate
the hadronic matrix elements. In the literature, many different approaches, such as
the $SU(6)$ quark model, bag models, QCD sum rules, and chiral perturbation theory have been
used to make estimations. In this subsection, we summarize the results
and get some idea about their uncertainties.

\subsubsection{Contribution from Quark EDM}

In the $SU(6)$ constituent quark model, the matrix elements of the
quark tensor operators are simple and scale-independent
\cite{He:1989xj,Beall:1981zq}, leading to
\begin{equation}
d_N^{(1)}=-\frac{1}{3}d^E_u+\frac{4}{3}d^E_d  \ .
\end{equation}
Although it has been suggested that one should use the constituent
quark masses in the formulas of quark EDM~\cite{Beall:1981zq}, this
is incorrect from the point of view of factorization.

In the parton quark model discussed in \cite{Abel:2004te}, it was found,
\begin{equation}
d_N^{(1)}=-0.508d^E_u+0.746d^E_d-0.226d^E_s \ .
\end{equation}
From the QCD sum rules, one gets~\cite{Pospelov:2000bw}
\begin{equation}
d_N^{(1)}=(1\pm0.5)\times0.7(-0.25d^E_u+d^E_d) \ .
\end{equation}
Different approximations are largely consistent.

\subsubsection{Contribution from Quark CDM}

The contribution to nEDM from the quark CDM in the constituent quark
model is~\cite{He:1989xj}
\begin{equation}\label{quarkcdm}
d_N^{(2)}=\frac{4}{9}\frac{e}{g_s}d^C_u+\frac{8}{9}\frac{e}{g_s}d^C_d
\ ,
\end{equation}
where $g_s$ is the coupling of strong interaction at the energy
scale where the model is applicable. In this calculation, the
authors assumed first that the neutron is composed of constituent
quarks, and then treated the gluon field inside the neutron as a
background, neglecting its kinetic energy. Therefore, Eq.
(\ref{quarkcdm}) can only be seen as an order-of-magnitude estimate.

Weinberg's naive dimensional analysis has also been used to estimate
this
contribution~\cite{Weinberg:1989dx,Arnowitt:1990eh,Arnowitt:1990je},
\begin{equation}
d_N^{(2)}\sim\frac{e}{4\pi}\left(O(1)d_u^C + O(1)d_d^C\right)\ .
\end{equation}
In Ref.~\cite{Khatsymovsky:1992yg}, the authors used the chiral
perturbation theory to calculate the singular part of the long
distance contribution,
\begin{equation}
d_N\simeq\frac{0.7e}{g_s}(d^C_u + d^C_d)\;.
\end{equation}
And finally, QCD sum rules analysis in Ref.~\cite{Pospelov:2000bw}
gives
\begin{equation}
d_N^{(2)}=(1\pm0.5)\times\frac{0.55e}{g_s}(0.5d_u^C+d_d^C) \ ,
\end{equation}
where $g_s$ is the strong coupling constant at 1 GeV, about 2.5.

\subsubsection{Contribution from Weinberg Operator}

The contribution from the Weinberg's operator $O_W$ can be estimated
by Weinberg's naive dimensional analysis~\cite{Weinberg:1989dx},
which is an order-of-magnitude estimate
\begin{equation}
d_N^{(3)}\simeq eMC_g(\mu)/4\pi\approx 100\;{\rm MeV}\; e\; C_g(1
{\rm GeV})\ ,
\end{equation}
where $M=4\pi F_\pi\simeq1190$~MeV and $\mu$ is the hadronic scale
taking as 1 GeV.

On the other hand, the estimate based on QCD sum rules gives \cite{Demir:2002gg}
\begin{equation}
d_N^{(3)}\simeq(10-30){\rm MeV} \;e\;C_g(1\;{\rm GeV}) \ ,
\end{equation}
which is considerably smaller. In any case, because of the small coefficient
function, the Weinberg operator contribution can essentially be neglected.

\subsubsection{Contribution from Four-Quark Operators}

The hadronic matrix elements of the four-quark operators have been
studied and reviewed in Ref.~\cite{An:2009zh}. In this work we will
take the results from that paper.

\subsection{Numerical Results}

As discussed in Ref.~\cite{Zhang:2007fn}, combining with the kaon
indirect CP-violation $\epsilon$ parameter, one can use nEDM to get
the most stringent lower bound on the mass of the right-handed $W$
boson in the context of the mLRSM. In Ref.~\cite{Zhang:2007fn}, the
authors used naive factorization~\cite{He:1992db} to estimate the
contribution of four-quark operators. However, this method for
baryons may not be valid even in the large-$N_C$ limit, and the
uncertainty is unknown. Therefore, we have assumed a very large
error on their matrix elements and the resulting constraint on the
left-right symmetry scale is not very strong. In a dedicated study
of these matrix elements~\cite{An:2009zh}, we have gotten a much
better understanding on their contribution. In
Ref.~\cite{An:2009zh}, the contribution of four-quark operators to
nEDM was separated into two parts, the direct contribution and the
meson-condensate contribution. For the direct contribution, quark
models were employed to calculate the hadronic matrix elements,
which is only an order-of-magnitude estimate be. However, for the
meson-condensate contribution, the factorization method was used to
calculate the meson matrix elements, which can be justified in the
large-$N_C$ limit. Since the meson-condensate contribution dominates
over the direct one, we believe that we reached a factor-of-two
accuracy in the matrix elements of four-quark operators.

\begin{figure}[hbt]
\begin{center}
\includegraphics[width=8cm]{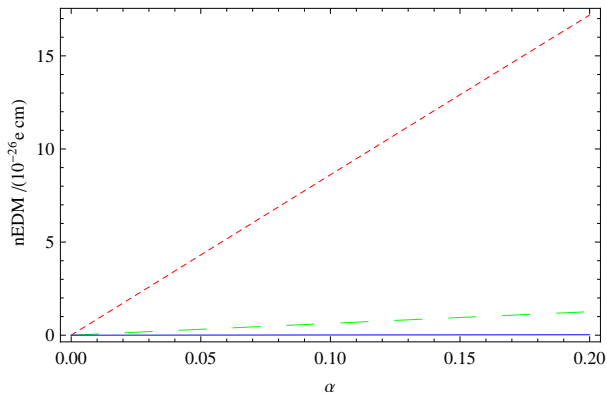}
\caption{nEDM contributed from operators, $\bar ui\gamma_5u\bar d d$
(short dashed red line), $\bar ui\gamma_5u\bar s s$ (long dashed
green line), down quark EDM and CDM operators (solid blue
line).}\label{alpha}
\end{center}
\end{figure}

In mLRSM, after neglecting the contributions from FCNH and the
charged higgs boson exchange, nEDM depends only on three parameters,
$r$, $\alpha$, and $M_{W_R}$, where $\alpha$ is the new source of
CP-violation. Therefore, if $\alpha=0$, nEDM predicted by the mLRSM
will be the same as that predicted by SM, about five orders of
magnitude smaller than the upper bound given by the current
experiment~\cite{Shabalin:1978rs}. Whereas for $\epsilon$, there are
two new contributions in mLRSM~\cite{Zhang:2007fn}, the Dirac phase
in the righthanded CKM matrix inherited from the lefthanded CKM
matrix, and the spontaneous phase $\alpha$. The new contribution
from the Dirac phase is enhanced compared to the similar
contribution in SM due to the chiral enhancement in the hadronic
matrix element (see Ref.~\cite{Bigi:2000yz} for a good review). The
contribution of the spontaneous CP-phase $\alpha$ must be adjusted
to cancel the contribution of the Dirac phase. Therefore, in mLRSM
there is a tension between nEDM and $\epsilon$ that one cannot only
adjust $\alpha$ to suppress all the new CP-violation sources, and a
large $M_{W_R}$ is needed. As a result, nEDM and $\epsilon$ together
give a lower bound on $M_{W_R}$.

\begin{figure}[hbt]
\begin{center}
\includegraphics[width=8cm]{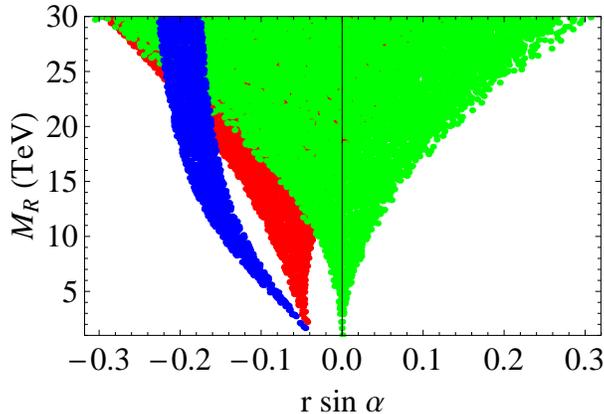}
\caption{Constraints on the mass of $W_R$ and the spontaneous
CP-violating parameter $\alpha$ from the kaon decay parameter
$\epsilon$ ($M_{H_0}=\infty$, red dots; $M_{H_0}=50$ TeV, blue dots)
and nEDM (green dots). For nEDM, we use the current experimental
upper bound as the constraint and for $\epsilon$ we use the criteria
that the beyond-SM-physics contribution should not exceed $1/4$ of
the experimental value. }\label{edmcon}
\end{center}
\end{figure}

In this new study, we use the QCD sum rules to estimate the
contribution of the quark EDM and CDM operators, and use the results
in Ref.~\cite{An:2009zh} for the contribution of the four-quark
operators. Fig.~\ref{alpha} shows the contributions to nEDM from
different operators at fixed $M_{W_R}$ and $r$. The result from the
Weinberg operator is too small to be included in the figure. It is
clear that the contributions from four-quark operators are much
larger than from quark EDM and CDM operators. One way to understand
this is that in mLRSM the quark EDM and CDM operators are generated
in the same way as the four-quark operators. The quark EDM and CDM
operators are generated through diagrams in Fig. \ref{soni} and the
four-quark operators are generated through diagrams in Fig.
\ref{fourquark}. The Wilson coefficients roughly have the following
relations
\begin{equation}
d^E_q \simeq \frac{em_qA}{16\pi^2}C_4\;;\;d^C_q \simeq
\frac{g_sm_qA'}{16\pi^2}C_4\;,
\end{equation}
where $A$ and $A'$ are two proportionality coefficients, $C_4$ is
the Wilson coefficient of certain four-quark operators. Take the
down quark EDM as an example, $A$ can be written as $\sin^2\theta_C
m_c/m_u\simeq 15$, where $\theta_C$ is the Cabibbo angle. From QCD
sum rules, nEDM contributed by the down-quark EDM operator is
approximately the down-quark EDM itself, whereas the nEDM contributed
directly from the four-quark operator can be written as
\cite{An:2009zh}
\begin{equation}
d_N^{\rm four-quark}\simeq\frac{e}{16\pi^2}B_0C_4\;,
\end{equation}
where $B_0 \simeq 2.2$ GeV is related to SSB of the chiral symmetry.
Since $B_0\gg A m_d$, nEDM directly from the four-quark operator
$\bar ui\gamma_5u \bar d d$ is much larger than the contribution
from the down quark EDM operator. Indeed, this is a common
phenomenon in left-right models and two-Higgs-doublet models, where
the quark EDM and CDM operators are always generated by the triangle
diagrams in Fig. \ref{soni}, and the internal lines are always
quarks. In other types of new physics models, the internal lines can
be other kind of fermions. For example, in supersymmetric models,
they can be gauginos, and in extra dimension models, they can be
KK-fermions, where the above relation between quark EDM operators
and four-quark operators is no longer hold. In these models, quark
EDM and CDM operators might be more important that four-quark
operators.

Using the matrix elements in Ref.~\cite{An:2009zh}, we calculate the
constraint from the nEDM and kaon-decay parameter $\epsilon$ on the
allowed parameter space of mLRSM. The result is shown in Fig.
\ref{edmcon}. The allowed parameter region by the experimental upper
bound on nEDM is shown as green dots. The constraints from
$\epsilon$-parameter depends strongly on the mass of the FCNH in the
theory. We have shown two possible values of $M_{H_0}$, 50 TeV and
$\infty$ for simplicity. We assume  for $\epsilon$ the new
contribution should not exceed 1/4 of the experimental value. From
Fig. \ref{edmcon} one can see that the lower bound for the $M_{W_R}$
from nEDM and $\epsilon$ is around 10 TeV. If we assume a factor of
2 uncertainty on the hadronic matrix elements, the actual bound is
$10\pm 3$ TeV. This will make a direct detection of the right-handed
gauge boson very difficult at LHC if it exits.

%Another CP-violation observable which also needs to be take into
%account is the direct CP-violation parameter in the $K-\bar K$
%system, the $\epsilon'$ parameter. Recently, $\epsilon'$ has been
%studied in the context of the mLRSM in
%Ref.~\cite{xxx}(Panying's paper). By setting the criteria that the
%new contribution should not exceed 1/4 of the experimental value, we
%can get the allowed parameter space which is shown in Fig.
%\ref{epsilonprime}.
%\begin{figure}[hbt]
%\begin{center}
%\includegraphics[height=5cm]{together1.eps}
%\includegraphics[height=5cm]{together2.eps}
%\caption{The yellow region is the parameter space of mLRSM
%constrained by $\epsilon'$ by setting the criteria that the new
%contribution not exceed 1/4 of the experimental value while the
%orange region is by setting that the new contribution not exceed the
%experimental value.}\label{epsilonprime}
%\end{center}
%\end{figure}

\section{conclusion}

In this paper, we have studied nEDM in mLRSM systematically by using
effective field theory approach. The formula for calculating nEDM is
given in Eq. (\ref{fac}). The contribution of four-quark operators
is found to be the most important. The contribution of Weinberg
operator to nEDM has been discussed systematically. A numerical
suppression is found which counteracts the infrared enhancement and
makes the contribution of this operator negligible. We have found a
lower bound on the mass of $W_R$ which is about $(10\pm3)$ TeV. This
is higher than what have been found before
\cite{Zhang:2007fn,Chen:2008kt} and certainly cannot be detected at
LHC.

In a more complicated non-supersymmetric scenario of LRSM, although
the CP-violation pattern in the Higgs sector might be change, the
tension between $\epsilon$ and nEDM discussed in Sec. IV still
exists. Therefore, one can also use this analysis to set a lower
bound on the righthanded scale. In the supersymmetric LRSM, there
are new CP-violation sources from the soft terms, which can
contribute to both nEDM and $\epsilon$. Furthermore, in
supersymmetric LRSM~\cite{Aulakh:1997ba}, the lefthanded and
righthanded CKM matrices must be equal to each other up to a sign,
therefore, if one assumes certain scenarios of the breaking
mechanism of supersymmetry, $\epsilon$ itself can give a constraint
on the righthanded scale~\cite{Zhang:2007qma}.

\acknowledgments

This work was partially supported by the U. S. Department of Energy
via grant DE-FG02-93ER-40762. F. Xu acknowledges a scholarship
support from China's Ministry of Education.

 %%%%%%%%%%%%%%%%%%%%%%%%%%%

%%%%%%%%%%%%%%%%%%%%%%%%%%%%%%%%%%%%%%%%%%%%%%%%%

\end{document}